\documentclass[twocolumn,preprintnumbers,amsmath,amssymb]{revtex4}

\usepackage{amsmath,amssymb,amsfonts} 	
\usepackage[dvips]{graphicx}
\usepackage[latin1]{inputenc}
\usepackage{subfigure}
\begin{document}

\title{Memory effects in non-adiabatic molecular dynamics at metal surfaces}

\author{Thomas Olsen}
\email{tolsen@fysik.dtu.dk}
\author{Jakob Schi{\o}tz}

\affiliation{Danish National Research Foundation's Center for Individual
Nanoparticle Functionality (CINF),
	Department of Physics,
	Technical University of Denmark,
	DK--2800 Kongens Lyngby,
	Denmark}

\date{\today}

\begin{abstract}
We study the effect of temporal correlation in a Langevin equation describing non-adiabatic dynamics at metal surfaces. For a harmonic oscillator the Langevin equation preserves the quantum dynamics exactly and it is demonstrated that memory effects are needed in order to conserve the ground state energy of the oscillator. We then compare the result of Langevin dynamics in a harmonic potential with a perturbative master equation approach and show that the Langevin equation gives a better description in the non-perturbative range of high temperatures and large friction. Unlike the master equation, this approach is readily extended to anharmonic potentials. Using density functional theory we calculate representative Langevin trajectories for associative desorption of N$_2$ from Ru(0001) and find that memory effects lowers the dissipation of energy. Finally, we propose an ab-initio scheme to calculate the temporal correlation function and dynamical friction within density functional theory.
\end{abstract}

\maketitle

\section{Introduction}
Modern computational surface chemistry, as e.g. applied to heterogeneous catalysis, is largely based on the Born-Oppenheimer approximation and potential energy surfaces which are typically obtained using density functional theory (DFT).\cite{christensen} In the adiabatic approximation the electrons are assumed to follow the motion of the nuclei instantaneously and the dynamics thus becomes confined to the ground state potential energy surface. While the adiabatic approximation has certainly been successful in giving a detailed quantitative account of a range of chemical reactions on metal surfaces, it is still not clear under which general circumstances the approximation is reliable.\cite{luntz1, juaristi1, luntz3, juaristi2} In particular, the role of non-adiabatic effects is often difficult to asses due to the inadequacy of low dimensional models of surface dynamics. For example, unusual sticking coefficients in the measured dissociative adsorption of N$_2$ on Ru(0001) hints at strong non-adiabatic energy loss,\cite{diekhoner1} but has been accounted for by multi-dimensional adiabatic dynamics.\cite{diaz1,diaz2} For other reactions, such as associative desorption of N$_2$ from Ru(0001), non-adiabatic effects still seem to be very important\cite{murphy, diekhoner2,luntz1} and multi-dimensional adiabatic simulations have not been able to account for large energy losses during desorption.\cite{diaz3}  Another example where adiabatic dynamics have failed is the dissociation of O$_2$ on Al(111) where spin selection rules gives rise to highly non-adiabatic behavior.\cite{behler1}

Non-adiabatic dynamics for isolated molecules is usually handled by including the first few excited adiabatic potential energy surfaces and imposing some surface hopping scheme. When distinct adsorbate diabatic states are present there may be physical arguments why the adsorbate should remain in such a state during a reaction and the non-adiabatic dynamics can be evaluated by constraining the adsorbate to such a diabat.\cite{behler2} This picture may then be improved by introducing surface hopping between diabats. However, for molecules adsorbed on metal surfaces there is an infinity of electronic excited states in the immediate vicinity of the ground state and surface hopping may not be the most practical scheme. Another popular and rather general method to handle non-adiabatic effects is through Langevin dynamics where electronic friction and stochastic forces account for dissipation and fluctuation as a result of coupling to excited electronic states.\cite{metiu, schmid, caldeira, brandbyge} Usually, the so-called Markov approximation is employed where the fluctuating forces are not temporally correlated but can be related to the electronic friction through the fluctuation-dissipation theorem.\cite{tully} At high electronic temperatures, thermal excitations dominate the electronic system and the Markov approximation is good for describing chemical reactions mediated by hot electrons.\cite{luntz2,olsen5} However, for non-adiabatic dynamics in general, the Markov approximation may fail and it then becomes important to take into account the 'memory' of the system. 

In the present paper we explore the consequences of Langevin dynamics with and without the Markov approximation at various temperatures. We follow the approach of Brandbyge et al. \cite{brandbyge} and base the analysis on a model Hamiltonian from which the electronic friction and correlation function can be derived explicitly. We start by modelling the internal stretch mode of CO adsorbed on Cu(100) by a considering a harmonic oscillator coupled to a thermal reservoir of electrons and compare the results to those obtained with a master equation approach. The general trend we see is that at low temperatures the Markov approximation overestimates the effect of dissipation. This is because the fluctuating forces are of thermal origin in the Markov approximation while the dissipative terms originate from non-thermal excitations and the relative effect of dissipation compared to fluctuations is thus increased. We also show that non-Markovian dynamics is needed in order to ensure energy conservation at low temperatures and thus maintains detailed balance between fluctuations and dissipation. Associative desorption of N$_2$ from Ru(0001) is then studied using the Langevin equation on representative trajectories and again, memory effects are shown to reduce the dissipation of energy. Finally, we comment on a possible method to obtain the full correlation function and thus include memory effects in an ab-initio setting based on DFT.\cite{trail} In appendix \ref{reduced}, we review the connection between the reduced density matrix and the Langevin equation and emphasize the probabilistic interpretation of the correlation function. In appendix \ref{discretization}, we review how correlated stochastic forces can be randomly sampled given a discretized version of the correlation function.

\section{Model}
A commonly used electronic Hamiltonian describing of atoms or molecules adsorbed on metals surfaces is the Newns-Anderson model,\cite{newnsanderson1,newnsanderson2} where the adsorbate is described by a single adsorbate state $|a\rangle$ which hybridizes with metallic states $|k\rangle$ and thus acquires a broadening in energy. A very simple non-adiabatic extension of this model is obtained by coupling the resonant states $|a\rangle$ to an adsorbate degree of freedom $x$ and extending the Hamiltonian with a nuclear kinetic energy and adiabatic potential. Assuming a quadratic nuclear potential and linear coupling to the resonance, the Hamiltonian becomes
\begin{align}\label{H}
H&=H_{el}+H_0+H_I,\\
H_0&=\frac{p^2}{2M}+\frac{1}{2}M\omega_0^2x^2,\notag\\
H_{el}&=\varepsilon_0c_a^\dag c_a+\sum_k\epsilon_k c_k^\dag c_k+\sum_k(V_{ak}c_a^\dag c_k+V_{ak}^*c_k^\dag c_a),\notag\\
H_I&=-fc_a^\dag c_ax\notag,
\end{align}
where $p$ is the nuclear momentum, $M$ the adsorbate effective mass, and $c_a^\dag$ and $c_k^\dag$ are creation operators for adsorbate and metallic electronic states respectively. The Hamiltonian $H_0+H_I$ can be thought of as a harmonic oscillator which is shifted when the resonance becomes occupied and the coupling constant $f$ is the force felt by adsorbate in this state. We will furthermore restrict ourselves to the wide band approximation in which the metallic band of electrons is assumed to be much wider than the width of the resonant state. The resonance projected density of states is then a Lorentzian:
\begin{align}\label{pdos}
\rho_a(\varepsilon)=\frac{1}{\pi}\frac{\Gamma/2}{(\varepsilon-\varepsilon_0)^2+(\Gamma/2)^2},
\end{align}
with full width at half maximum given by
\begin{align}
\Gamma=2\pi\sum_k|V_{ak}|^2\delta(\varepsilon_0-\epsilon_k).
\end{align}

Due to the non-adiabatic coupling in Eq.~\ref{H}, the adsorbate may exchange energy with the electronic system via the resonant state $|a\rangle$. However, usually we are only interested in the nuclear degrees of freedom and it is then convenient to trace out the electronic degrees of freedom from the full dynamics. This is accomplished by the reduced time dependent density matrix:
\begin{align}\label{density_matrix}
\rho_{red}(t)=\text{Tr}_{el}\Big(e^{-iHt/\hbar}\rho_0e^{iHt/\hbar}\Big),
\end{align}
where $\text{Tr}_{el}$ means the trace over electronic states and $\rho_0$ is the full density matrix at $t=t_0$. Choosing an adsorbate basis $|\nu\rangle$, the diagonal elements of the reduced density matrix give the probabilities that the adsorbate is in a particular state at time $t$.

\subsection{Non-Markovian master equation}
We will consider the time-dependent probability of being in a particular energy eigenstate $|n\rangle$. The equation governing these probabilities is known as a master equation and can be derived by taking the trace of the Liouville equation for the full density matrix. The result is 
\begin{align}\label{liouville}
\frac{d\rho_{red}}{dt}+\frac{i}{\hbar}[H_0,\rho_{red}]=-i\mathcal{F}[\rho],
\end{align}
where the influence functional $\mathcal{F}[\rho]=\text{Tr}_{el}[H_I,\rho]/\hbar$ depends on the complete history of the full density matrix. Gao \cite{gao97} has shown how to evaluate $\mathcal{F}[\rho]$ using the Hamiltonian \eqref{H} within the self-consistent Born approximation. Furthermore, imposing the Markov approximation, where it is assumed that the influence functional only depends on the instantaneous value of the density matrix, and taking the diagonal elements of Eq.~\eqref{liouville} led to an explicit expression for the master equation. However, using the formalism of Gao \cite{gao97}, it is straightforward to generalize the results to a non-Markovian master equation. For completeness we state the result here which is
\begin{align}\label{master}
\dot{p}_m(t)=2f^2\sum_n|x_{mn}|^2&\int_{t_0}^tdt'\Big[\widetilde{W}_{n\rightarrow m}(t-t')p_n(t')\\
&-\widetilde{W}_{m\rightarrow n}(t-t')p_m(t')\Big],\notag
\end{align}
with the differential rates given by
\begin{align}\label{rate}
\widetilde{W}_{n\rightarrow m}(t)=&\int_{-\infty}^\infty d\omega_1\int_{-\infty}^\infty d\omega_2\rho_a(\omega_2)\big(1-n_F(\omega_2)\big)\\
&\times\rho_a(\omega_1)n_F(\omega_1)\cos[(\omega_1-\omega_2+\omega_{nm})t],\notag
\end{align}
where $p_m(t)=\langle m|\rho_{red}(t)|m\rangle$, $|m\rangle$ is an eigenstate of $H_0$ with eigenvalue $\varepsilon_m$, $x_{mn}=\langle m|x|n\rangle$, $n_F(\varepsilon)$ is the Fermi-Dirac distribution, $\rho_a(\varepsilon)$ is the projected density of states \eqref{pdos} and $\omega_{nm}=(\varepsilon_n-\varepsilon_m)/\hbar$. 

The Markov approximation is obtained by extending the temporal integration to infinity which is justified when $t$ is much larger than some electronic correlation time $t_c$. The probabilities $p_m$ are then assumed to depend on t rather than $t'$ and integrating over $t'$ yields the usual golden rule type expression for the transition rates.\cite{gao97} The master equation was derived assuming that off-diagonal elements of the density matrix are not important. Allthough it is straightforward to generalize the result \eqref{master} to a coherent master equation which takes off-diagonal elements into account, it has been shown that, if the initial state is not coherent, the off-diagonal elements will have very little influence on the diagonal elements.\cite{gao97} In terms of multiple inelastic scattering events, neglecting coherency corresponds to associating a probability distribution to each scattering event.\cite{olsen1,olsen3}

\begin{figure}[t]
	  \includegraphics[width=8.5 cm, clip]{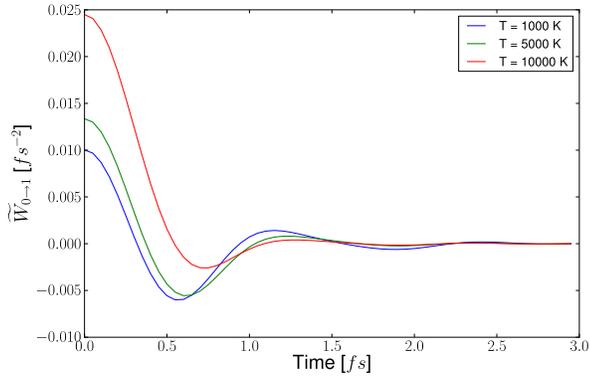}
\caption{The differential transition rate $W_{0\rightarrow1}(t)$ given in Eq.~\eqref{rate} using three different temperatures and $\Gamma=2.0\;eV$, $\varepsilon_0=2.6\;eV$, $\hbar\omega_0=0.250\;eV$ and $f=8.7\;eV/$\AA. The figure shows that the time dependence vanishes after a few femtoseconds and since the typical timescale of change in $p_n(t)$ is $\sim100\;fs$, the Markov approximation is expected to work well for the master equation.}
\label{fig:W}
\end{figure}

In Fig.~\ref{fig:W} we show the differential transition rate $\widetilde{W}_{0\rightarrow1}(t)$ at three different temperatures. The time dependence only depends on the properties of the electronic system at the given temperature and the non-adiabatic coupling $f$ simply gives an overall scaling. Since the probabilities $p_n(t)$ typically change on timescales of $\sim100\;fs$ and the $\widetilde{W}_{n\rightarrow m}(t)$ approach zero within a few femtoseconds, the Markov approximation is expected to be very good for the master equation in a large range of temperatures.

\subsection{Non-Markovian Langevin dynamics}
If we calculate the diagonal of the reduced density matrix in a basis of position eigenstates, a Langevin equation emerges. This is achieved by writing Eq.~\eqref{density_matrix} as a path integral and using the Feynman-Vernon formalism of influence functionals to obtain an effective action to second order in the frictional coupling $f$.\cite{feynman-vernon, caldeira, schmid} The result is given in Eq.~\eqref{density_integrated}. This approach is not perturbative in the same sense as the master equation where the derivation is based on a direct second order expansion of the reduced density matrix. Rather, the second order expansion of the action leads to a density matrix which contains all orders of the frictional coupling.
As explained in appendix \ref{reduced}, the result can be interpreted as a sum over classical Langevin trajectories with initial conditions sampled from the Wigner distribution of the initial state and the equation of motion is
\begin{align}\label{langevin}
M\ddot{x}(t)+M\omega^2x(t)+\int_{t_0}^tdt'\eta(t-t')\dot{x}(t')=\xi(t),
\end{align}
where $\eta(t)$ the dynamical electronic friction and $\xi(t)$ is a Gaussian distributed stochastic force specified by its correlation function: 
\begin{align}
\langle\xi(t)\xi(t')\rangle=K(t-t').
\end{align}

With the model Hamiltonian \eqref{H} it is possible to evaluate the friction and correlation function explicitly. The result is:\cite{brandbyge}
\begin{align}\label{friction}
\eta(t)=\int_{-\infty}^\infty\frac{d\omega}{2\pi}\Lambda(\omega)\cos(\omega t),
\end{align}
with
\begin{align}\label{lambda}
\Lambda(\omega)=\frac{\hbar}{\omega}&\int_{-\infty}^\infty\frac{d\omega_1}{2\pi}\int_{-\infty}^\infty d\omega_2G(\omega_1,\omega_2)\\
&\times\delta\Big(\omega-(\omega_2-\omega_1)\Big)\Big(n_F(\omega_1)-n_F(\omega_2)\Big),\notag
\end{align}
\begin{align}\label{gamma}
G(\omega_1,\omega_2)=4\pi^2f^2\rho_a(\omega_1)\rho_a(\omega_2),
\end{align}
and $\rho_a(\omega)$ is the projected density of states Eq.~\eqref{pdos}. The correlation function is
\begin{align}\label{correlation_function}
K(t)=\frac{\hbar}{2}\int_{-\infty}^{\infty}\frac{d\omega}{2\pi}\omega\Lambda(\omega)\coth\Big(\frac{\hbar\omega}{2k_BT}\Big)\cos(\omega t).
\end{align}

\begin{figure}[t]
	  \includegraphics[width=8.5 cm, clip]{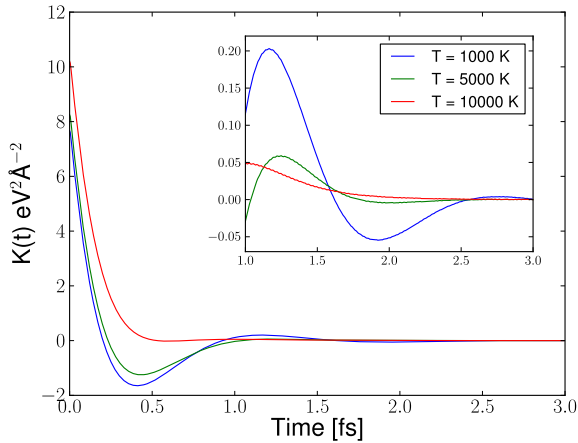}
\caption{The correlation function $K(t)$ given in Eq.~\eqref{correlation_function} using three different temperatures with $\Gamma=2.0\;eV$, $\varepsilon_0=2.6\;eV$, $\hbar\omega_0=0.250\;eV$ and $f=8.7\;eV/$\AA. At low temperatures correlation can persists for several femtoseconds.}
\label{fig:K}
\end{figure}
\begin{figure}[t]
	  \includegraphics[width=7.0 cm, clip]{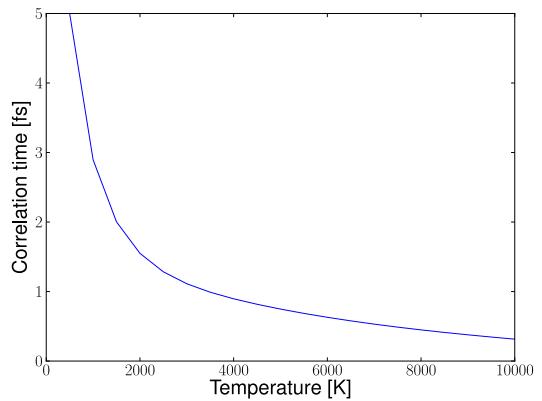}
\caption{The correlation time Eq.~\eqref{t_c} as a function of temperature with. Below $3500\;K$ the correlation time becomes larger than2 a femtosecond which is the largest time step we can use in the molecular dynamics and non-Markovian processes therefore begins to influence the dynamics below this temperature.}
\label{fig:tc}
\end{figure}

In Fig.~\ref{fig:K} we show the correlation function for three different temperature. The structure and typical correlation time is very similar to the differential rate shown in Fig.~\ref{fig:W}. However, in contrast to the master equation the timescale of motion in the Langevin equation is on the order of $\Delta t\sim1\;fs$ and correlation effects may be very important at low temperatures. In general the stochastic forces at a given time depends on the state of the electronic system which again depends on the path taken by the adsorbate. This gives rise to correlation between forces at different times and this ''memory'' is the price one pays for tracing out the electronic degrees of freedom. The range of memory in the system depends strongly on temperature since high temperatures tend to rapidly destroy coherence in the state of the electronic system. In the high temperature limit where $k_BT\gg\hbar/\Delta t$, one obtains the well known expression
\begin{align}\label{markov}
K(t)=2k_BT\eta_0\delta(t),
\end{align}
where $\eta_0=\Lambda(0)/2$. This is the Markov approximation in which there is no correlation between forces at different times. Taking CO on Cu(100) as an example, the smallest timescale is the period of vibrational motion which is $\sim16\;fs$. With a standard Verlet integration one needs a timestep of $\Delta t\sim 1fs$ to describe ground state vibrations and a first estimate of the validity of the Markov approximation is obtained as: $T\gg\hbar/(\Delta tk_B)\sim 2900\;K$. To get a better quantitative estimate of the validity of the Markov approximation we can consider the correlation time $t_c$ given by
\begin{align}\label{t_c}
t_c^2=\frac{\int dtt^2K(t)}{\int dtK(t)}.
\end{align}
It should be noted that the correlation time is only a function of the electronic system and does not depend on the non-adiabatic coupling $f$. We have calculated $t_c$ as a function of temperature and the result is shown in Fig.~\ref{fig:tc}. For molecular dynamics requiring a time step no larger than $\sim1\;fs$, we see that the correlation time becomes larger than this when the temperature comes below $3500\;K$. Thus below this temperature non-Markovian processes play an important role in the dynamics.

\section{Results}
Before we test the role of non-Markovian effects on a generic non-adiabatic surface reaction, we will compare Markovian and non-Markovian Langevin dynamics for a harmonic oscillator potential with results obtained from a master equations approach.

\subsection{Quadratic potential}\label{quadratic}
\begin{figure}[t]
	  \includegraphics[width=4.25 cm, clip]{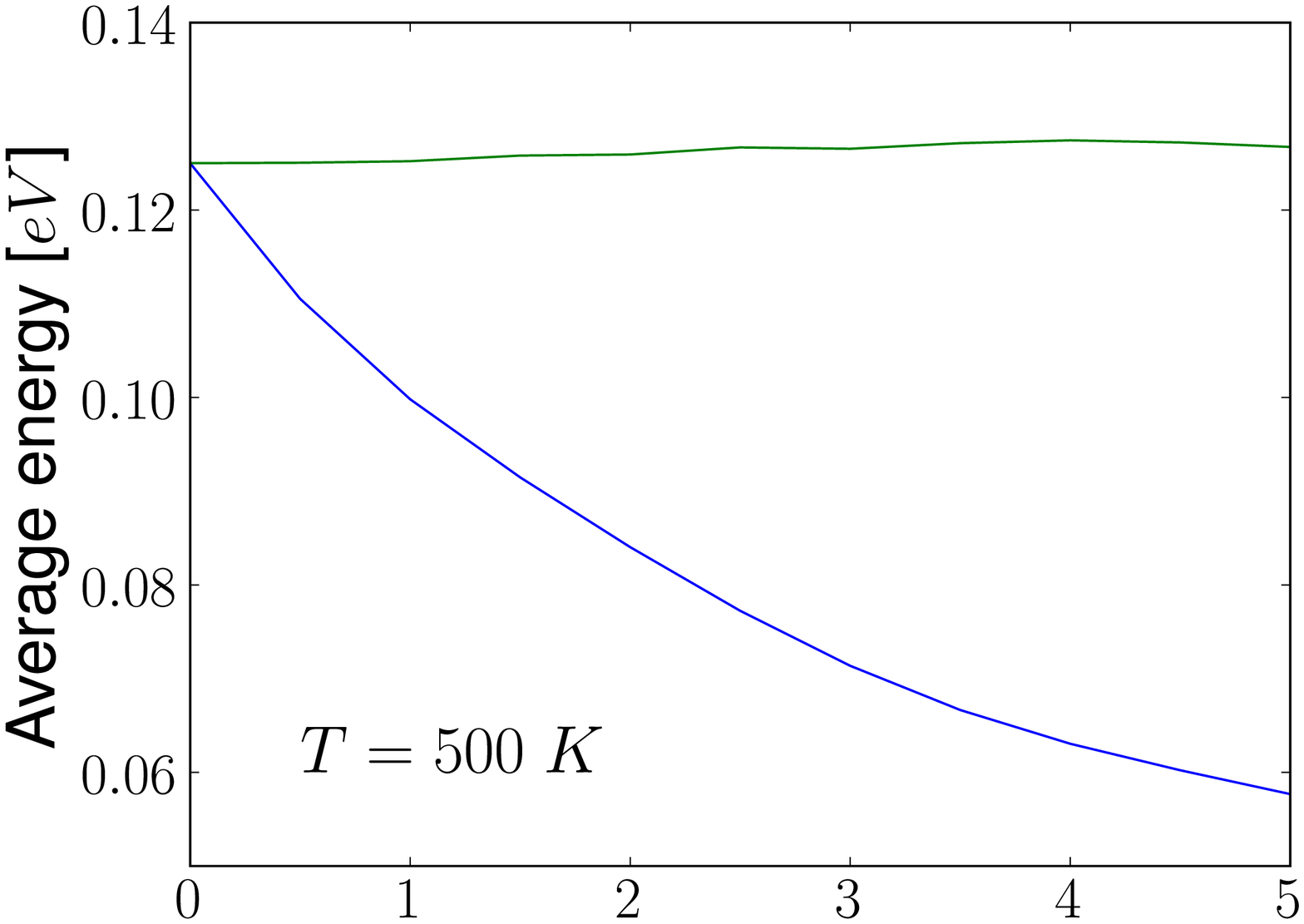}
	  \includegraphics[width=4.25 cm, clip]{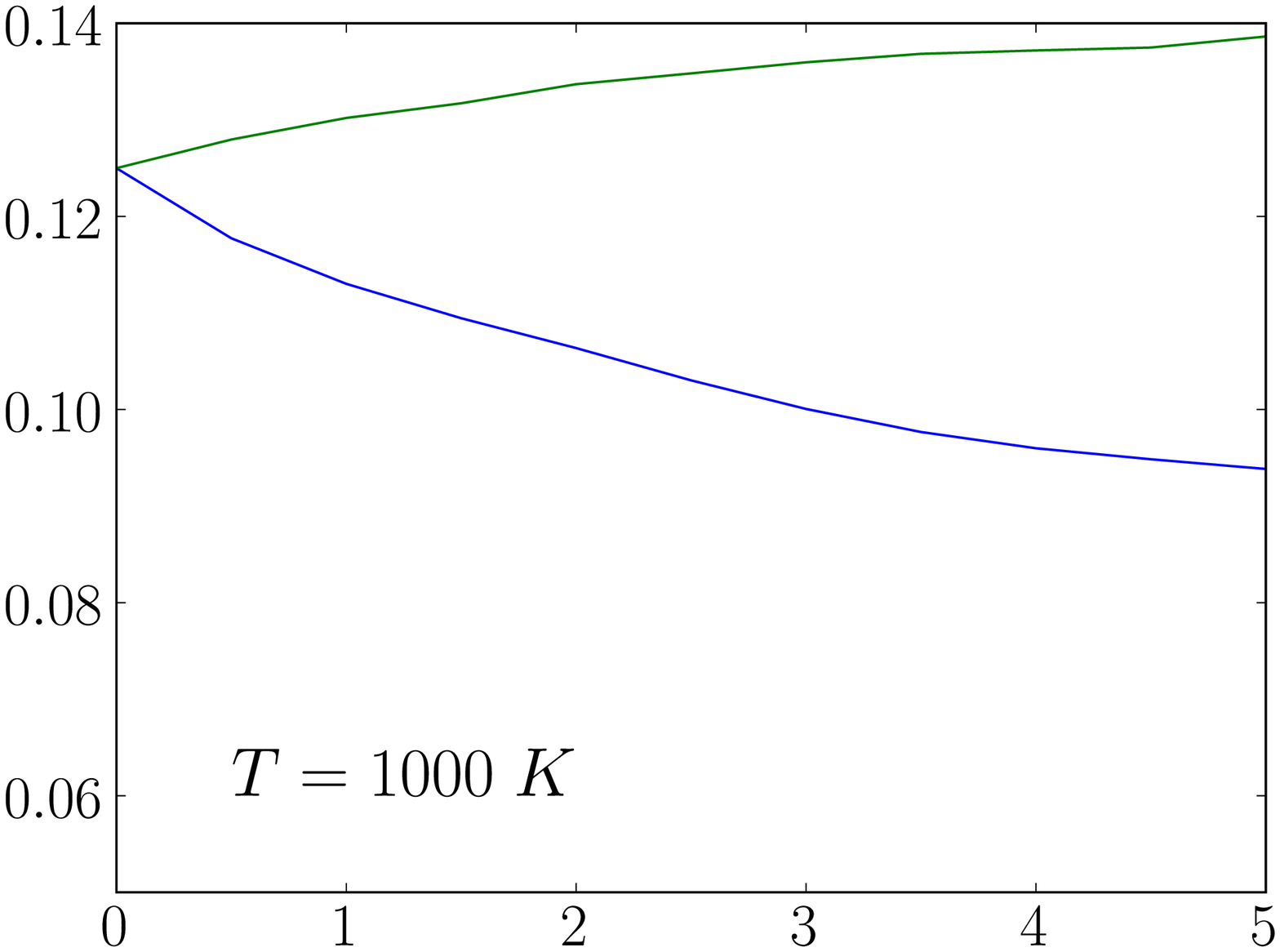}\\
	  \includegraphics[width=4.25 cm, clip]{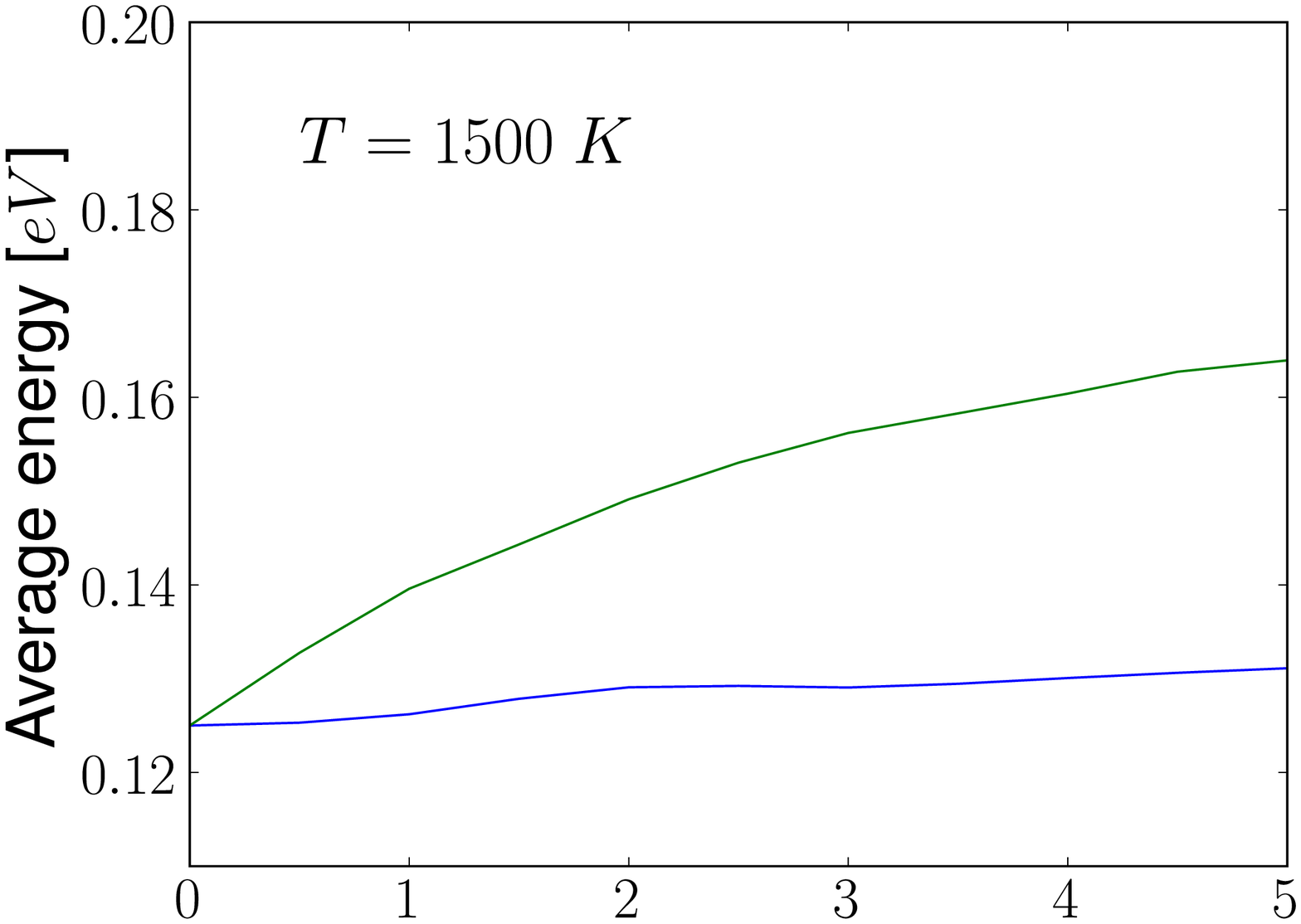}
	  \includegraphics[width=4.25 cm, clip]{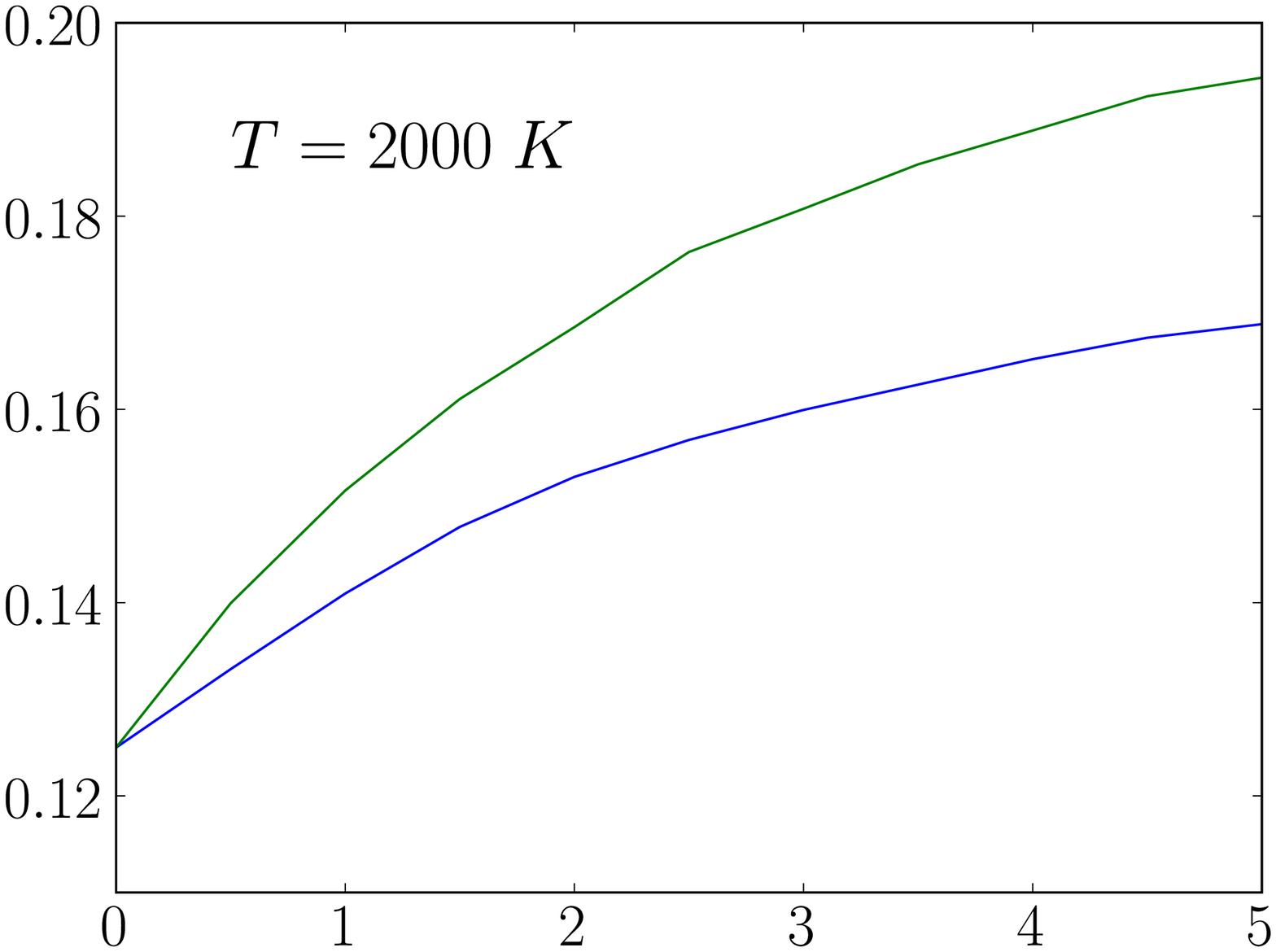}\\
	  \includegraphics[width=4.25 cm, clip]{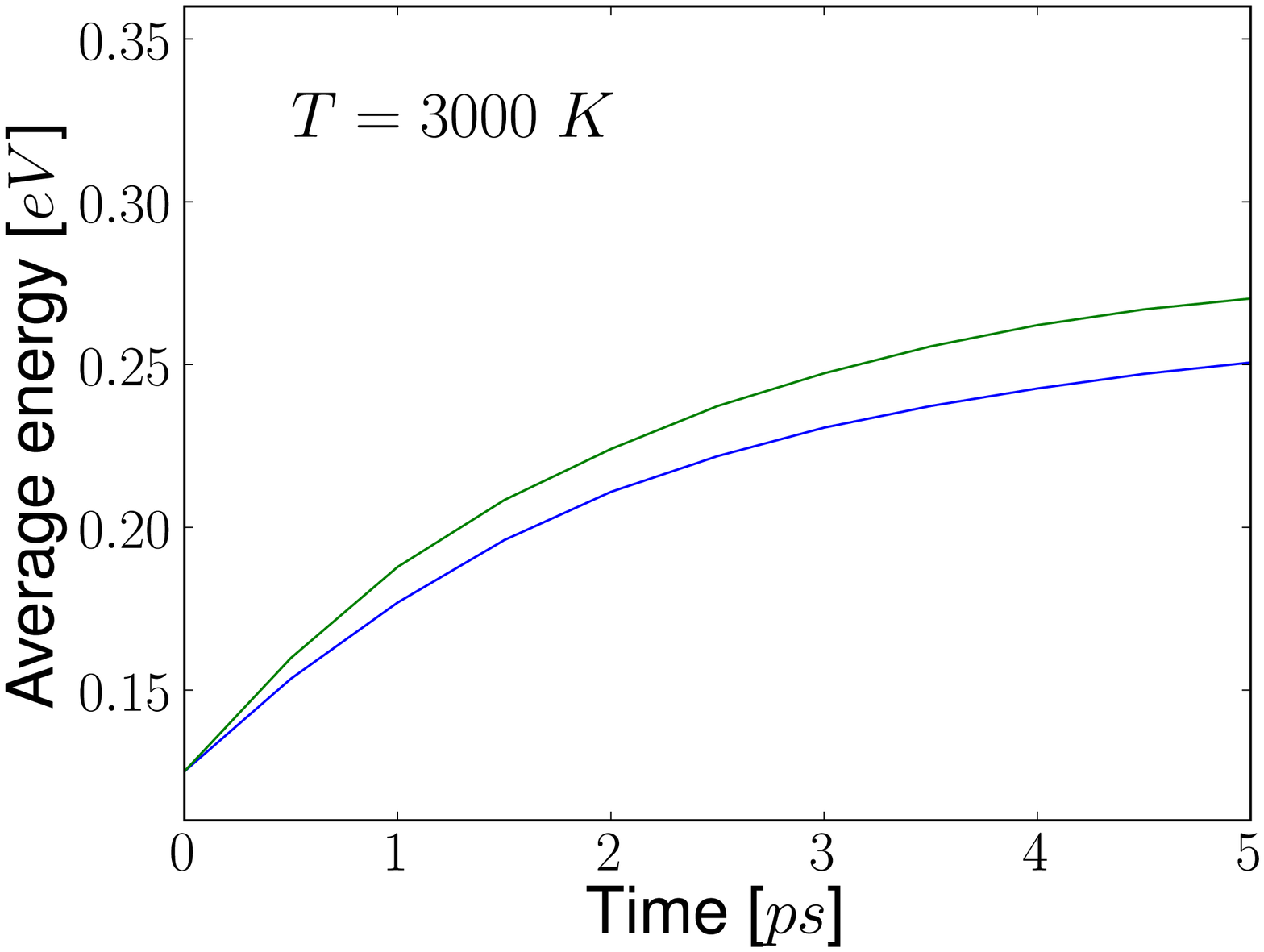}
	  \includegraphics[width=4.25 cm, clip]{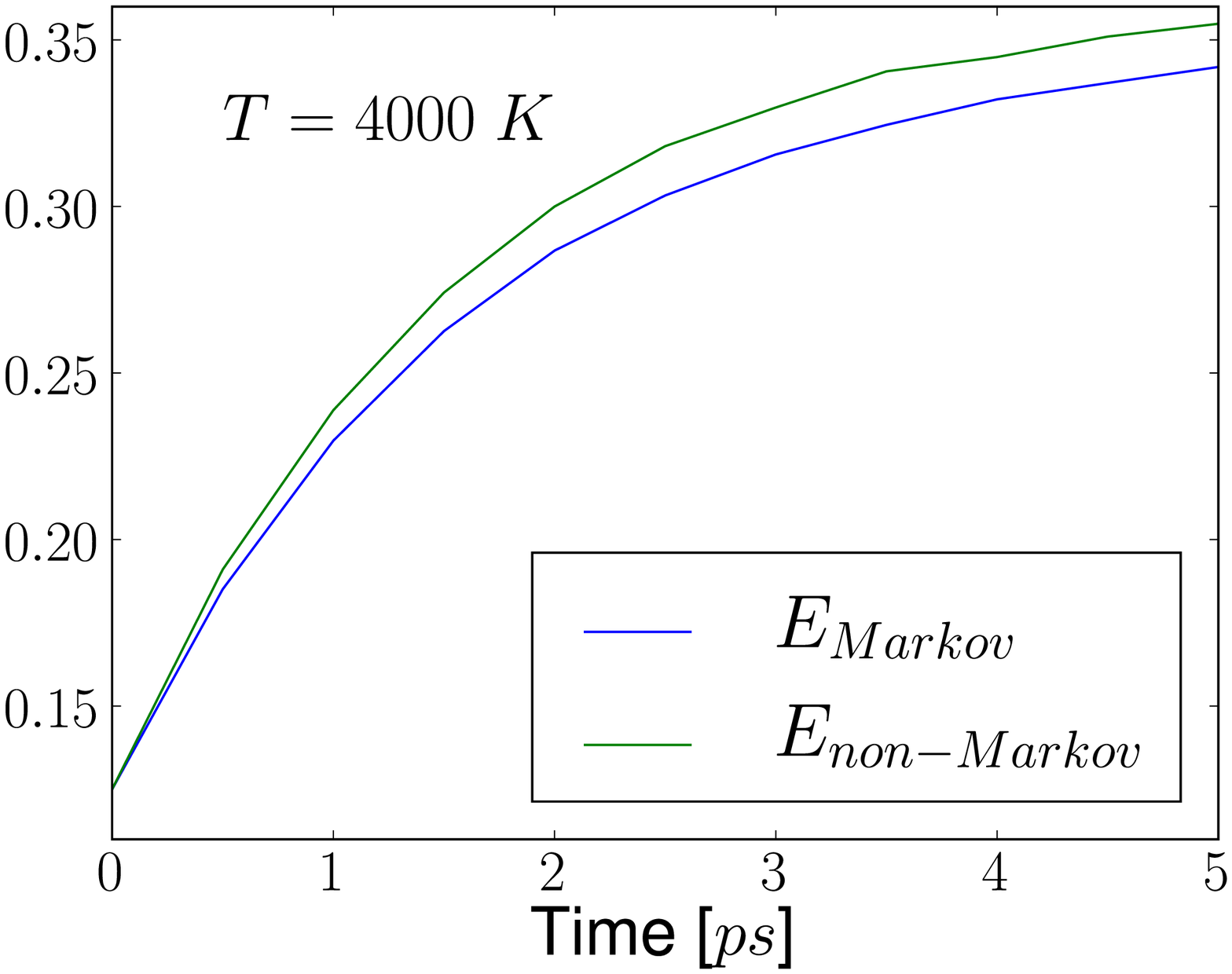}
\caption{Average energy of a harmonic oscillator interacting with a thermal reservoir of electrons at six different temperatures evaluated using Langevin dynamics with memory and with the Markov approximation. The Markov approximation fails below $T=3000\;K$ where quantum fluctuations are important.}
\label{fig:E_t}
\end{figure}

It is easy to see that the fluctuating force in the Langevin equation has to vanish within the Markov approximation Eq.~\eqref{markov} when $T_{el}\rightarrow0$. With a quadratic potential and $\eta(t)=\Lambda(0)/2\delta(t)$ it is then possible to solve the Langevin equation analytically which gives the time-dependent energy
\begin{equation}
 E_{Markov}(t)=E_0e^{-t/\tau},\qquad\tau=2M/\Lambda(0),
\end{equation}
where $E_0$ is the initial energy. However, as shown in Ref. \onlinecite{olsen5}, the Langevin equation is quantum mechanically exact for a harmonic potential if the initial conditions are accounted for correctly and the total energy should thus not be allowed to decay below $\hbar\omega/2$. The problem is that the Markov approximation neglects all non-thermal excitations of the electronic system and leads to pure dissipation at $T_{el}\rightarrow0$. In reality, an oscillating adsorbate will induce (non-thermal) excitations of the electron gas which may then influence the propagation of the adsorbate. In general, it is therefore expected that the Markov approximation tends to underestimate the influence of the electronic system on the adsorbate. This non-Markovian effect should vanish at high temperature where the thermal excitations of the electronic system dominate. In Fig.~\ref{fig:E_t} we show the time evolution of the average energy of a harmonic oscillator interaction with a thermal reservoir of electrons at six different temperatures. The average energy is calculated using the full non-Markovian correlation function as described in appendix \ref{discretization} and within the Markov approximation. The initial state was chosen as the vibrational ground state and included exactly by phase space sampling the Wigner distribution.\cite{olsen5} The parameters used were chosen to match the internal vibrational mode of CO adsorbed on Cu(100)\cite{olsen2,olsen5} and we have thus taken $\Gamma=2.0\;eV$, $\varepsilon_0=2.6\;eV$, $\hbar\omega=0.25\;eV$, and $f=-8.7\;eV/$\AA. The failure of the Markov approximation and resulting decay of the average energy is clearly seen at low temperatures. In particular, at $T=500\;K$ the Markov approximation gives rise to exponentially decaying energy whereas the energy remains nearly fixed at $E\approx E_0$ when memory effects are included. For high temperatures ($T>3000\;K$) thermal excitations dominate and the Markov approximation becomes reliable. In all calculations we have converged the results by decreasing the time steps.

\begin{figure}[t]
	  \includegraphics[width=8.5 cm, clip]{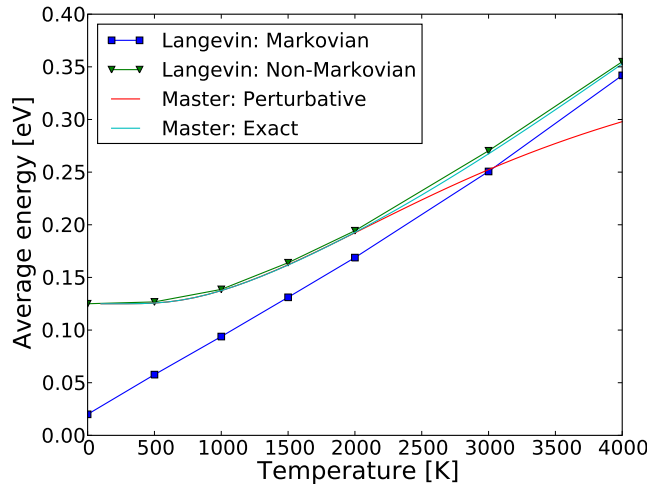}
\caption{Average energy of a harmonic oscillator after $5\;ps$ of interaction with a thermal reservoir of hot electrons. The non-Markovian Langevin equation and the non-perturbative master equation both give the correct dependence, whereas the Markovian Langevin equation fails at low temperatures and the perturbative master equation fails at high temperatures.}
\label{fig:E_T}
\end{figure}

In Fig.~\ref{fig:E_T} we show the average energy of the harmonic oscillator after $5\;ps$ of interaction with a thermal reservoir of electrons. The energy is calculated with non-Markovian Langevin dynamics, Markovian Langevin dynamics, the master equation with rates obtained from perturbation theory and the master equation with exact (non-perturbative) rates.\cite{gao97} Using the full non-Markovian master equation Eqs.~\eqref{master}-\eqref{rate} does not change the results.
The non-Markovian Langevin approach matches the exact non-perturbative Master equation approach, whereas the perturbative Master equation fails at high temperatures and the Markovian Langevin approach fails at low temperatures.
It may be surprising that the non-Markovian Langevin equation reproduces the exact and not the perturbative master equation. However, as mentioned above, the perturbative derivation of the master equation is based on a direct evaluation of the reduced density matrix to second order in the non-adiabatic coupling,\cite{gao97} while the Langevin equation is derived by constructing an effective action to second order in the non-adiabatic coupling.\cite{brandbyge} Thus, while the reduced density matrix calculated from the effective action only becomes exact in the small friction limit, it does contain terms to all orders in the non-adiabatic coupling and is a much better approximation for large frictional coupling and high temperatures than the direct perturbative derivation leading to the master equation Eqs.~\eqref{master}-\eqref{rate}.

It may seem like a complete overkill to apply a non-adiabatic Langevin dynamics to a harmonic potential when the results are readily obtainable from the master equation approach. However, for anharmonic potentials it is not possible to derive transition rates for the master equation exactly and the best approximation is then the non-Markovian Langevin equation. This was also concluded in Ref. \onlinecite{olsen5} where a perturbative master equation approach underestimated transition rates in a Morse potential compared to a Markovian Langevin approach.

\subsection{Associative desorption of N$_2$ from Ru(0001)}
As an example of a potential where the master equation approach is not readily applicable, we consider the well-known example of associative desorption of N$_2$ from Ru(0001). This system has been subject to extensive experimental\cite{diekhoner1, diekhoner2, murphy} and theoretical\cite{murphy,diaz1,diaz2,diaz3,luntz1} studies and much evidence points to a non-adiabatic dissipation of energy during associative desorption.

The Langevin equation can be generalized to an arbitrary potential $V_0(x)$ by a semiclassical expansion of the potential and the excited state forces acting on the adsorbate.\cite{brandbyge} The potential is included by making the substitution $M\omega_0^2x^2/2\rightarrow V_0(x)$ in the Hamiltonian Eq.~\eqref{H} and the friction arises from an excited resonant state with potential energy $V_1(x)$ and is included by the generalizations $-fx\rightarrow\varepsilon_a(x)=V_1(x)-V_0(x)$ and $V_{ak}\rightarrow V_{ak}(x)$ which in the wide band limit leads to a position dependent resonance width $\Gamma(x)$. When multiple coordinates $\{x_i\}$ are considered $\eta$, $\Lambda$, and $K(t)$ in Eqs.~\eqref{friction}-\eqref{correlation_function} become tensors and the Langevin equations for each coordinate become coupled through terms like $\sum_j\eta_{ij}(x)\dot{x}_j$. In addition to temporal correlation, the stochastic forces acting on different coordinates become correlated through the off-diagonal terms in the correlated function:
\begin{align}\label{spatial}
\langle\xi_i(t)\xi_j(0)\rangle=K_{ij}(t).
\end{align}
In fact, since the friction tensor is well approximated by $\Lambda_{ij}(\omega)\propto f_if_j$ with $f_i=\partial\varepsilon_a(x)/\partial x_i$, $K_{ij}(t)$ has only one non-zero eigenvalue. This implies that there is a single (coordinate dependent) mode on which the stochastic force acts and the random forces can thus be regarded as completely spatially correlated at any given time.

\begin{figure}[t]
	  \includegraphics[width=8.5 cm, clip]{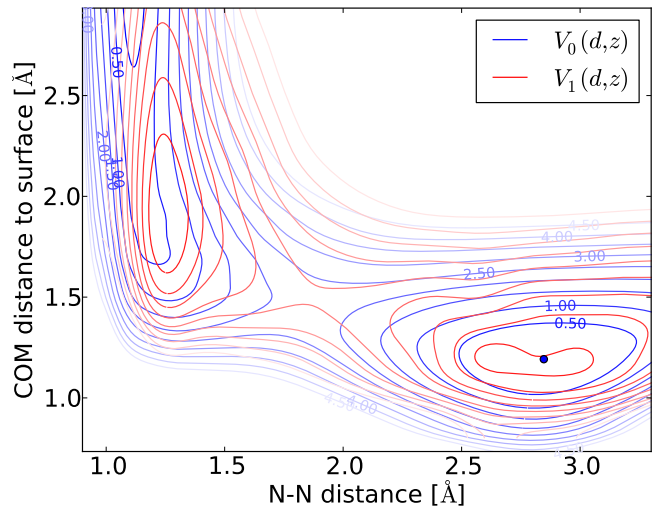}
\caption{Ground and excited state potential energy surfaces of N$_2$ adsorbed on Ru(0001). The excited state was obtained by occupying the $2\pi$ orbital of N$_2$.}
\label{fig:pes}
\end{figure}

We have studied associative desorption of N$_2$ from Ru(0001) using the code \texttt{gpaw},\cite{gpaw, mortensen} which is a real-space Density Functional Theory (DFT) code that uses the projector augmented wave method.\cite{blochl1,blochl2} The Ru(0001) substrate was modelled by a three layer slab where the top layer was relaxed. We used a grid spacing of $0.2$ {\AA} and the calculations were performed in a (2x4) supercell sampled by a (4x6) grid of $k$-points using the RPBE\cite{hammer} exchange-correlation functional. The friction is assumed to be dominated by the $2\pi$ orbital which is only partly occupied in the ground state. To calculate the excited state potential energy $V_1(x)$ we applied a generalization of the $\Delta$-self-consistent field method where the resonant state is expanded in a basis of Kohn-Sham orbitals and the resulting resonant density is added to the density in each iteration step. Thus for each adsorbate position we calculate the energy resulting from forcing an electron into a $2\pi$ orbital. For details on the method and comparison with experiments we refer to Ref. \onlinecite{gavnholt} We have restricted the analysis to the two-dimensional desorption process considered in Ref. \onlinecite{murphy} where the two N atoms are adsorbed at adjacent hcp hollow sites and desorbs by moving perpendicular to the bridge towards the fcc hollow while changing the center of mass coordinate. While a two-dimensional analysis is almost certainly not sufficient to obtain quantitative results,\cite{luntz1, diaz2,diaz3} we do expect to draw some qualitative conclusions about the validity of the Markov approximation for this system. The calculated ground and excited state potential energy surfaces are shown in Fig.~\ref{fig:pes}. To obtain $\Gamma(d,z)$ we have fitted the width of the projected density of states of the $2\pi$ orbital along the minimum reaction to an exponential $\Gamma(z)=\Gamma_0e^{-(z-z_t)/z_d}$ and obtained $\Gamma_0=3.0\;eV$, $z_d=0.5$ {\AA} and $z_t$ is the center of mass position at the transition state. The frictional force $\partial\varepsilon_a(d,z)/\partial d$ in the internal mode become large in the exit channel and gives rise to large dissipation of the internal energy while the molecule desorbs. However due to the rapid decay of $\Gamma(z)$ the friction tensor essentially vanishes at $z=3$ {\AA}. The amount of dissipated energy thus largely depends on the time spend in the exit channel in the immediate vicinity of the transition state.

To examine the impact of non-adiabatic dissipation of energy and, in particular, the validity of the Markov approximation, we have considered four representative initial conditions leading to desorption. All four are initially at the transition state with a kinetic energy of $0.1\;eV$. The kinetic energy is then concentrated in positive or negative center of mass momentum or positive or negative internal momentum. We have taken the surface and thus the electronic temperature to be $T=900\;K$.\cite{murphy,luntz1}. Table \ref{tab1} displays the average energy loss in a desorption event of the four initial conditions with and without the Markov approximation. The reason for the large difference is due to the average time spend in the exit channel which for initial negative internal momentum is $\sim125\;fs$ and for initial positive center of mass momentum is $\sim250\;fs$. The shift to lower dissipation in non-Markovian dynamics is what we would expect from the conclusions in Sec. \ref{quadratic} and Fig \ref{fig:E_t}. In general, memory effects tends to increase the importance of fluctuating forces and thus decrease the overall dissipation of energy. 

The present analysis should in no way be taken as a quantitative study of non-adiabatic effect in associative desorption of N$_2$ from Ru(0001). In such a study one would need to sample a thermal distribution of initial configurations at the transition state and include all 6 molecular degrees of freedom.\cite{diaz3} Furthermore, the present study assumes that the electronic friction originates from a single resonance ($2\pi$) and it is well described by the wide band approximation. That this is not the case has already been established\cite{luntz1} and a complete ab-initio scheme for the electronic friction is needed. Such a scheme based on DFT has already been suggested and put to use within the Markov approximation\cite{trail, luntz1, luntz2}, but need to be generalized slightly to take memory effects into account. In section \ref{DFT} we will propose such a generalization.
\begin{table}[t]
\begin{center}
\begin{tabular}{c|c|c|c|c}
Mode & z- & z+ & d- & d+ \\
\hline
Markovian     & 0.31 & 0.49 & 0.079 & 0.10\\
non-Markovian & 0.13 & 0.42 & 0.053 & 0.10\\
\end{tabular}
\end{center}
\caption{Average energy loss (all numbers in eV) of trajectories leading to desorption for four different initial conditions with and without the Markov approximation at $T=900\;K$. The initial conditions were all at the transition state with a kinetic energy of $0.1\;eV$. $d$ and $z$ denotes initial momentum in the internal vibrational mode and the center of mass mode respectively and - and + denotes the sign of the initial momentum. In general, the Markov approximation tends to underestimate the effect of fluctuating forces which results in too much dissipation.}
\label{tab1}
\end{table}

For anharmonic potentials where the friction tensor acquires a position dependence, non-Markovian Langevin dynamics is rather time consuming, since one has to calculate and diagonalize the correlation function in each time step. In the present simulation, the time required for a single time step in the dynamics was increased by a factor of $10^3$ compared to Markovian dynamics, but the computational time is, however, vanishing compared to that required for a full DFT calculation at a given position. For N$_2$ on Ru(0001) the memory effects are clearly seen but probably not important compared to neglecting four degrees of freedom. The calculated energy dissipation is not large enough to account for the vibrational damping observed in Ref. \onlinecite{diekhoner2}, but it is very likely that inclusion of more degrees of freedom would result in a larger amount of time spend in the exit channel and thus a much larger dissipation of internal energy.
 
\section{Non-Markovian friction and fluctuating forces from density functional theory}\label{DFT}
Within linear response theory, it is possible to derive an expression for the electronic friction for a general non-adiabatic Hamiltonian if one assumes classical adsorbate motion.\cite{trail} The result depends on the response function
of the electronic system as well as the derivative of the electron-vibron coupling with respect to adsorbate coordinates. 
Replacing the true response function with a Kohn-Sham response function and the coupling Hamiltonian by a Kohn-Sham potential, give the result for the electronic friction
\begin{align}\label{friction_dft}
\eta=-\pi\hbar\sum_{ij}\big|\langle\psi_i|&\frac{dV_{KS}}{dx}|\psi_j\rangle\big|^2\\
&\times\int d\varepsilon\frac{dn_F(\varepsilon)}{d\varepsilon}\delta(\varepsilon_i-\varepsilon)
\delta(\varepsilon_j-\varepsilon),\notag
\end{align}
where $\psi_i$ are Kohn-Sham orbitals with eigenenergies $\varepsilon_i$. The result is valid within the Markov approximation, since the memory in the Kohn-Sham potential has been neglected. Generalizing this result to include non-Markovian dynamics would require a non-adiabatic exchange-correlation potential, which is presently out of reach. However, since the result is equivalent to that obtained within the reduced density matrix formalism and the Hamiltonian \eqref{H} we can impose a very simple generalization which reduces to the adiabatic result \eqref{friction_dft} in the Markov approximation and to 
\eqref{friction}-\eqref{correlation_function} in the case of non-interacting electrons. Indeed, it is easy to verify that \eqref{friction_dft} reduces to the Markovian limit of \eqref{friction}-\eqref{correlation_function} if $V_{KS}$ is replaced with the Hamiltonian \eqref{H} and one is led to a non-Markovian Langevin equation based on DFT which is given by \eqref{langevin}-\eqref{lambda} and \eqref{correlation_function}, but with
\begin{align}\label{G_dft}
G(x;\omega_1,\omega_2)=4\pi^2\sum_{ij}&\big|\langle\psi_i|\frac{dV_{KS}}{dx}|\psi_j\rangle\big|^2\\
&\times\delta(\omega_1-\varepsilon_i/\hbar)\delta(\omega_2-\varepsilon_j/\hbar)\notag.
\end{align}
This result for the dynamical friction was also derived in Ref. \onlinecite{trail}, however, the dominating non-Markovian effect is the correlation function \eqref{correlation_function} which follows from the reduced density matrix formalism in conjunction with \eqref{G_dft}.

\section{Summary}
From a fundamental point of view it is important to realize that the Langevin equation gives an exact description of a harmonic oscillator interacting with a reservoir of electrons if the initial quantum state is taken correctly into account.\cite{olsen5} However, it is easy to see that the fluctuating forces vanish at low temperatures in the Markov approximation, which then results in a exponentially decaying energy of the oscillator. This of course contradicts the quantum description of the oscillator and the problem can be traced to the Markov approximation which does not take non-thermal electronic excitations into account. In Fig.~\ref{fig:E_t}, we have shown explicitly how memory effects 'saves' the quantum behavior of the oscillator and conserves the energy of the vibrational ground state at low electronic temperatures.

Another way of handling dissipative systems is using the master equation. This approach is based on a perturbative calculation of the reduced density matrix in basis of energy eigenstates. While this approach is fast and intuitively appealing, the method breaks down at high temperature or large friction due to the perturbative nature of the method. In contrast, the Langevin equation is based on an effective action Eq.~\eqref{rho_red}-\eqref{S_eff} giving a non-perturbative flavor. Furthermore, the master equation requires quantization of the potential energy surface and becomes impractical for complicated potentials with many bound states.

As an example of such a potential we have considered the associative desorption of N$_2$ from Ru(0001). We have not tried to perform a quantitative two-dimensional study of this system as done by Luntz et al.\cite{luntz1}, but rather examined the effect of temporal correlation in two representative trajectories. As expected, the effect is an increased significance of the fluctuating forces leading to lower dissipation when memory is included. While this is may be of qualitative interest, the effect of including all degrees of freedom and performing an ab-initio calculation of the friction tensor, would almost certainly lead to corrections which are quantitatively much more important.\cite{diaz3}

Finally, we have provided an expression for the correlation function within an ab-initio DFT scheme. The result follows naturally by combining the usual DFT based friction tensor\cite{trail} with the relationship between the friction and fluctuating forces in a non-adiabatic Newns-Anderson Model.\cite{brandbyge} In principle, this scheme allows one to model non-adiabatic dynamics at metal surfaces by Langevin dynamics with ab-initio non-Markovian friction and fluctuating forces. 

\begin{acknowledgments}
This work was supported by the Danish Center for Scientific Computing. The Center for Individual Nanoparticle Functionality (CINF) is sponsored by the Danish National Research Foundation.
\end{acknowledgments}

\appendix
\section{Path integral representation of the reduced density matrix}\label{reduced}
In this appendix we give a path integral representation of the reduced density matrix Eq.~\eqref{density_matrix} in a coordinate basis. We will focus on the probabilistic interpretation of the path integral which leads to a Gaussian distributed classical Langevin equation. 

In a coordinate representation the reduced density matrix is 
\begin{align}
\rho_{red}(x,y;t)=\langle x|Tr_{el}[\rho(t)]|y\rangle.
\end{align}
and the diagonal elements give the probabilities of finding the adsorbate at a particular position regardless of the state of the electronic system. As shown in Ref. [\onlinecite{brandbyge}] the reduced density matrix of the Hamiltonian \eqref{H} can be represented as a double path integral:
\begin{widetext}
\begin{align}\label{rho_red}
\rho_{red}(x,y;t)=\int dx_0dy_0\langle x_0|\rho_0|y_0\rangle\int\mathcal{D}[x(t')]\mathcal{D}[y(t')] e^{iS_{eff}[x(t'),y(t')]/\hbar},
\end{align}
with the effective action given by
\begin{align}\label{S_eff}
S_{eff}[x(t'),y(t')]=S_0[x(t')]-S_0[y(t')]&-\int_0^tdt'\int_0^{t'}dt''v(t')\eta(t'-t'')\dot{u}(t'')\\
&+\frac{i}{2\hbar}\int_0^tdt'\int_0^tdt''v(t')K(t'-t'')v(t'')\notag,
\end{align}
where $u(t)=x(t)/2+y(t)/2$, $v(t)=x(t)-y(t)$ and $\eta(t)$ and $K(t)$ are given by Eqs.~\eqref{friction} and \eqref{correlation_function} respectively. With a quadratic potential the non-interacting action is given by $S_0[x(t')]=M\int_{t_0}^tdt'(\dot{x}^2-\omega_0^2x^2)/2$. Changing to coordinates $u$ and $v$ and performing a partial integration on the kinetic term then gives for the diagonal part of the density matrix:
\begin{align}\label{density_final}
\rho_{red}(u;t)=\int du_0dp_0\mathcal{P}(u_0,p_0)\int\mathcal{D}[u(t')]\mathcal{D}[v(t')] e^{-\frac{i}{\hbar}\int_0^tdt'\xi(t')v(t')-\frac{1}{2\hbar^2}\int_0^tdt'dt''v(t')K(t'-t'')v(t'')},
\end{align}
where $\mathcal{P}(x_0,p_0)$ is the Wigner distribution of $\rho_0$,
\begin{align}\label{xi}
\xi(t)=M\ddot{u}(t)+M\omega_0^2u^2(t)+\int_{t_0}^tdt'\eta(t-t')\dot{u}(t'),
\end{align}
and $u(t')$ have the additional constraint that $\dot{u}_0=p_0/m$. For a non-quadratic potential $V(x)$, one is forced to make a semiclassical expansion of the potential to second order and the exponential in Eq.~\eqref{density_final} would contain additional terms of order $\mathcal{O}(v^3V''')$. Similarly, with a nonlinear interaction $H_I$ in \eqref{H} one can perform a semiclassical expansion of the frictional terms which leads to a position dependence in Eq.~\eqref{gamma}. 

Without the quadratic term in $v(t')$, the density matrix \eqref{density_final} would give a delta functional in $\xi(t)$ and the dynamics would be governed by a classical equation of motion with dynamical friction function $\eta(t)$. However the last term in the exponential of \eqref{density_final} gives rise to a Gaussian broadening of the classical path. To see this explicitly we "complete" the square in the exponential and perform the path integral in $v(t')$ which gives
\begin{align}\label{density_integrated}
\rho_{red}(u;t)\propto\int du_0dp_0\mathcal{P}(u_0,p_0)\int\mathcal{D}[u(t')] e^{-\frac{1}{2}\int_0^tdt'dt''\xi(t')K^{-1}(t'-t'')\xi(t'')},
\end{align}
where $K^{-1}$ solves
\begin{align}
\int_0^tdt''K^{-1}(t'-t'')K(t''-t''')=\delta(t'-t''').
\end{align}
The exponential in \eqref{density_integrated} can be interpreted as the probability density of taking the path $u(t')$ given the endpoints $u_0$ and $u(t)$ and the initial velocity $\dot{u}_0=p_0/m$. It has a maximum at $\xi(t)=0$ corresponding to the classical dynamics and the classical path is broadened by $K^{-1}$. However, it will be more convenient to consider the probability density of $\xi(t)$ which obviously has dimensions of a force. It is then necessary to change the path integral measure from $\mathcal{D}[u(t')]$ to $\mathcal{D}[\xi(t')]$ and it can be shown that the Jacobian of this transformation is independent of $u(t')$.\cite{schmid} The two-point correlation function of $\xi(t)$ can then be calculated by 
\begin{align}
\langle\xi(t_1)\xi(t_2)\rangle&=\frac{\int\mathcal{D}[\xi(t')]\xi(t_1)\xi(t_2) e^{-\frac{1}{2}\int_0^tdt'dt''\xi(t')K^{-1}(t'-t'')\xi(t'')}}{\int\mathcal{D}[\xi(t')] e^{-\frac{1}{2}\int_0^tdt'dt''\xi(t')K^{-1}(t'-t'')\xi(t'')}}\notag\\
&=\frac{\frac{\delta^2}{\delta J(t_1)\delta J(t_2)}\int\mathcal{D}[\xi(t')] e^{-\frac{1}{2}\int_0^tdt'dt''\xi(t')K^{-1}(t'-t'')\xi(t'')-\int_0^tdt'J(t')\xi(t')}\bigg|_{J=0}}{\int\mathcal{D}[\xi(t')] e^{-\frac{1}{2}\int_0^tdt'dt''\xi(t')K^{-1}(t'-t'')\xi(t'')}}\notag\\
&=\frac{\delta^2}{\delta J(t_1)\delta J(t_2)}e^{\frac{1}{2}\int_0^tdt'dt''J(t')K(t'-t'')J(t'')}\bigg|_{J=0}=K(t_1-t_2).
\end{align}
\end{widetext}
This is the most compact way of specifying the statistical properties of $\xi(t)$ and Eq.~\eqref{xi} can be regarded as a classical equation motion with a stochastic Gaussian distributed force $\xi(t)$.

\section{Discretization of the correlation function}\label{discretization}
To sample a correlated ''force path'' $\xi(t)$ we need to discretize the correlation function and diagonalize the resulting correlation matrix. For a set of Gaussian distributed stochastic variables $\{\xi_i\}$ with probability distribution
\begin{align}
 P(\{\xi_i\})\sim \exp\Big(-\frac{1}{2}\sum_{ij}\xi_iC_{ij}^{-1}\xi_j\Big).
\end{align}
The correlation matrix $C_{ij}$ can be assumed symmetric without loss of generality. Hence, there exist a diagonal basis of uncorrelated variables $\{\xi'_i\}$ which can be sampled from independent normalized Gaussians. The transformation can be obtained by a Cholesky decomposition of $C_{ij}$ such that $\xi_i=\sum_jL_{ij}\xi'_j$, where $\sum_jL_{ij}L_{kj}=C_{ik}$.

The stochastic force appearing in the Langevin equation can be regarded as an infinite number of stochastic variables; one for each point in time from $t_0$ to $t$. Thus, to obtain an expression for the fluctuation force in a time interval $\Delta t$, we need the statistical properties of the integrals
\begin{align}
 \xi_i=\frac{1}{\Delta t}\int_{i\Delta t}^{(i+1)\Delta t}\xi(t')dt'.
\end{align}
From the theory of multivariate Gaussian distributions it is readily shown that the set of these integrals are Gaussian distributed with the correlation matrix:
\begin{align}
 C_{ij}=\frac{1}{\Delta t^2}\int_{i\Delta t}^{(i+1)\Delta t}dt'\int_{j\Delta t}^{(j+1)\Delta t}dt''K(t'-t''),
\end{align}
and this is the expression used when calculating molecular trajectories using non-Markovian Langevin dynamics.


\begin{thebibliography}{10}%
\makeatletter
\providecommand \@ifxundefined [1]{%
 \ifx #1\undefined \expandafter \@firstoftwo
 \else \expandafter \@secondoftwo
\fi
}%
\providecommand \@ifnum [1]{%
 \ifnum #1\expandafter \@firstoftwo
 \else \expandafter \@secondoftwo
\fi
}%
\providecommand \enquote [1]{``#1''}%
\providecommand \bibnamefont  [1]{#1}%
\providecommand \bibfnamefont [1]{#1}%
\providecommand \citenamefont [1]{#1}%
\providecommand\href[0]{\@sanitize\@href}%
\providecommand\@href[1]{\endgroup\@@startlink{#1}\endgroup\@@href}%
\providecommand\@@href[1]{#1\@@endlink}%
\providecommand \@sanitize [0]{\begingroup\catcode`\&12\catcode`\#12\relax}%
\@ifxundefined \pdfoutput {\@firstoftwo}{%
 \@ifnum{\z@=\pdfoutput}{\@firstoftwo}{\@secondoftwo}%
}{%
 \providecommand\@@startlink[1]{\leavevmode}%
 \providecommand\@@endlink[0]{}%
}{%
 \providecommand\@@startlink[1]{%
  \leavevmode
  \pdfstartlink
   attr{/Border[0 0 1 ]/H/I/C[0 1 1]}%
   user{/Subtype/Link/A<</Type/Action/S/URI/URI(#1)>>}%
  \relax
 }%
 \providecommand\@@endlink[0]{\pdfendlink}%
}%
\providecommand \url  [0]{\begingroup\@sanitize \@url }%
\providecommand \@url [1]{\endgroup\@href {#1}{\urlprefix}}%
\providecommand \urlprefix [0]{URL }%
\providecommand \Eprint[0]{\href }%
\@ifxundefined \urlstyle {%
  \providecommand \doi [1]{doi:\discretionary{}{}{}#1}%
}{%
  \providecommand \doi [0]{doi:\discretionary{}{}{}\begingroup
  \urlstyle{rm}\Url }%
}%
\providecommand \doibase [0]{http://dx.doi.org/}%
\providecommand \Doi[1]{\href{\doibase#1}}%
\providecommand \selectlanguage [0]{\@gobble}%
\providecommand \bibinfo [0]{\@secondoftwo}%
\providecommand \bibfield [0]{\@secondoftwo}%
\providecommand \translation [1]{[#1]}%
\providecommand \BibitemOpen[0]{}%
\providecommand \bibitemStop [0]{}%
\providecommand \bibitemNoStop [0]{.\EOS\space}%
\providecommand \EOS [0]{\spacefactor3000\relax}%
\providecommand \BibitemShut [1]{\csname bibitem#1\endcsname}%
\bibitem{christensen}%
  \BibitemOpen
  \bibfield{author}{%
  \bibinfo {author} {\bibfnamefont{C.~H.}\ \bibnamefont{Christensen}}\ and\
  \bibinfo {author} {\bibfnamefont{J.~K.}\ \bibnamefont{N{\o}rskov}},\ }%
  \bibfield{journal}{%
  \bibinfo {journal} {J. Chem. Phys.}\ }%
  \textbf{\bibinfo {volume} {128}},\ \bibinfo {pages} {182503} (\bibinfo {year}
  {2008})\BibitemShut{NoStop}%
\bibitem{luntz1}%
  \BibitemOpen
  \bibfield{author}{%
  \bibinfo {author} {\bibfnamefont{A.~C.}\ \bibnamefont{Luntz}}\ and\ \bibinfo
  {author} {\bibfnamefont{M.}~\bibnamefont{Persson}},\ }%
  \bibfield{journal}{%
  \bibinfo {journal} {J. Chem. Phys.}\ }%
  \textbf{\bibinfo {volume} {123}},\ \bibinfo {pages} {074704} (\bibinfo {year}
  {2005})\BibitemShut{NoStop}%
\bibitem{juaristi1}%
  \BibitemOpen
  \bibfield{author}{%
  \bibinfo {author} {\bibfnamefont{J.~I.}\ \bibnamefont{Juaristi}}, \bibinfo
  {author} {\bibfnamefont{M.}~\bibnamefont{Alducin}}, \bibinfo {author}
  {\bibfnamefont{R.~D.}\ \bibnamefont{Mui\~no}}, \bibinfo {author}
  {\bibfnamefont{H.~F.}\ \bibnamefont{Busnengo}},\ and\ \bibinfo {author}
  {\bibfnamefont{A.}~\bibnamefont{Salin}},\ }%
  \bibfield{journal}{%
  \Doi{10.1103/PhysRevLett.100.116102}{\bibinfo {journal} {Phys. Rev. Lett.}}\
  }%
  \textbf{\bibinfo {volume} {100}},\ \bibinfo {pages} {116102} (\bibinfo
  {month} {Mar}\ \bibinfo {year} {2008})\BibitemShut{NoStop}%
\bibitem{luntz3}%
  \BibitemOpen
  \bibfield{author}{%
  \bibinfo {author} {\bibfnamefont{A.~C.}\ \bibnamefont{Luntz}}, \bibinfo
  {author} {\bibfnamefont{I.}~\bibnamefont{Makkonen}}, \bibinfo {author}
  {\bibfnamefont{M.}~\bibnamefont{Persson}}, \bibinfo {author}
  {\bibfnamefont{S.}~\bibnamefont{Holloway}}, \bibinfo {author}
  {\bibfnamefont{D.~M.}\ \bibnamefont{Bird}},\ and\ \bibinfo {author}
  {\bibfnamefont{M.~S.}\ \bibnamefont{Mizielinski}},\ }%
  \bibfield{journal}{%
  \Doi{10.1103/PhysRevLett.102.109601}{\bibinfo {journal} {Phys. Rev. Lett.}}\
  }%
  \textbf{\bibinfo {volume} {102}},\ \bibinfo {pages} {109601} (\bibinfo
  {month} {Mar}\ \bibinfo {year} {2009})\BibitemShut{NoStop}%
\bibitem{juaristi2}%
  \BibitemOpen
  \bibfield{author}{%
  \bibinfo {author} {\bibfnamefont{J.~I.}\ \bibnamefont{Juaristi}}, \bibinfo
  {author} {\bibfnamefont{M.}~\bibnamefont{Alducin}}, \bibinfo {author}
  {\bibfnamefont{R.~D.}\ \bibnamefont{Mui\~no}}, \bibinfo {author}
  {\bibfnamefont{H.~F.}\ \bibnamefont{Busnengo}},\ and\ \bibinfo {author}
  {\bibfnamefont{A.}~\bibnamefont{Salin}},\ }%
  \bibfield{journal}{%
  \Doi{10.1103/PhysRevLett.102.109602}{\bibinfo {journal} {Phys. Rev. Lett.}}\
  }%
  \textbf{\bibinfo {volume} {102}},\ \bibinfo {pages} {109602} (\bibinfo
  {month} {Mar}\ \bibinfo {year} {2009})\BibitemShut{NoStop}%
\bibitem{diekhoner1}%
  \BibitemOpen
  \bibfield{author}{%
  \bibinfo {author} {\bibfnamefont{L.}~\bibnamefont{Diekh{\"o}ner}}, \bibinfo
  {author} {\bibfnamefont{H.}~\bibnamefont{Mortensen}}, \bibinfo {author}
  {\bibfnamefont{A.}~\bibnamefont{Baurichter}},\ and\ \bibinfo {author}
  {\bibfnamefont{A.~C.}\ \bibnamefont{Luntz}},\ }%
  \bibfield{journal}{%
  \bibinfo {journal} {J. Chem. Phys.}\ }%
  \textbf{\bibinfo {volume} {115}},\ \bibinfo {pages} {3356} (\bibinfo {year}
  {2001})\BibitemShut{NoStop}%
\bibitem{diaz1}%
  \BibitemOpen
  \bibfield{author}{%
  \bibinfo {author} {\bibfnamefont{C.}~\bibnamefont{D\'\i{}az}}, \bibinfo
  {author} {\bibfnamefont{J.~K.}\ \bibnamefont{Vincent}}, \bibinfo {author}
  {\bibfnamefont{G.~P.}\ \bibnamefont{Krishnamohan}}, \bibinfo {author}
  {\bibfnamefont{R.~A.}\ \bibnamefont{Olsen}}, \bibinfo {author}
  {\bibfnamefont{G.~J.}\ \bibnamefont{Kroes}}, \bibinfo {author}
  {\bibfnamefont{K.}~\bibnamefont{Honkala}},\ and\ \bibinfo {author}
  {\bibfnamefont{J.~K.}\ \bibnamefont{N\o{}rskov}},\ }%
  \bibfield{journal}{%
  \bibinfo {journal} {Phys. Rev. Lett.}\ }%
  \textbf{\bibinfo {volume} {96}},\ \bibinfo {pages} {096102} (\bibinfo {year}
  {2006})\BibitemShut{NoStop}%
\bibitem{diaz2}%
  \BibitemOpen
  \bibfield{author}{%
  \bibinfo {author} {\bibfnamefont{C.}~\bibnamefont{D\'\i{}az}}, \bibinfo
  {author} {\bibfnamefont{J.~K.}\ \bibnamefont{Vincent}}, \bibinfo {author}
  {\bibfnamefont{G.~P.}\ \bibnamefont{Krishnamohan}}, \bibinfo {author}
  {\bibfnamefont{R.~A.}\ \bibnamefont{Olsen}}, \bibinfo {author}
  {\bibfnamefont{G.~J.}\ \bibnamefont{Kroes}}, \bibinfo {author}
  {\bibfnamefont{K.}~\bibnamefont{Honkala}},\ and\ \bibinfo {author}
  {\bibfnamefont{J.~K.}\ \bibnamefont{N\o{}rskov}},\ }%
  \bibfield{journal}{%
  \bibinfo {journal} {J. Chem. Phys.}\ }%
  \textbf{\bibinfo {volume} {125}},\ \bibinfo {pages} {114706} (\bibinfo {year}
  {2006})\BibitemShut{NoStop}%
\bibitem{murphy}%
  \BibitemOpen
  \bibfield{author}{%
  \bibinfo {author} {\bibfnamefont{M.~J.}\ \bibnamefont{Murphy}}, \bibinfo
  {author} {\bibfnamefont{J.~F.}\ \bibnamefont{Skelly}}, \bibinfo {author}
  {\bibfnamefont{A.}~\bibnamefont{Hodgson}},\ and\ \bibinfo {author}
  {\bibfnamefont{B.}~\bibnamefont{Hammer}},\ }%
  \bibfield{journal}{%
  \bibinfo {journal} {J. Chem. Phys.}\ }%
  \textbf{\bibinfo {volume} {110}},\ \bibinfo {pages} {6954} (\bibinfo {year}
  {1999})\BibitemShut{NoStop}%
\bibitem{diekhoner2}%
  \BibitemOpen
  \bibfield{author}{%
  \bibinfo {author} {\bibfnamefont{L.}~\bibnamefont{Diekh{\"o}ner}}, \bibinfo
  {author} {\bibfnamefont{L.}~\bibnamefont{Horne{\ae}r}}, \bibinfo {author}
  {\bibfnamefont{H.}~\bibnamefont{Mortensen}}, \bibinfo {author}
  {\bibfnamefont{E.}~\bibnamefont{Jensen}}, \bibinfo {author}
  {\bibfnamefont{A.}~\bibnamefont{Baurichter}}, \bibinfo {author}
  {\bibfnamefont{V.~V.}\ \bibnamefont{Petrunin}},\ and\ \bibinfo {author}
  {\bibfnamefont{A.~C.}\ \bibnamefont{Luntz}},\ }%
  \bibfield{journal}{%
  \bibinfo {journal} {J. Chem. Phys.}\ }%
  \textbf{\bibinfo {volume} {117}},\ \bibinfo {pages} {5018} (\bibinfo {year}
  {2002})\BibitemShut{NoStop}%
\bibitem{diaz3}%
  \BibitemOpen
  \bibfield{author}{%
  \bibinfo {author} {\bibfnamefont{C.}~\bibnamefont{D\'\i{}az}}, \bibinfo
  {author} {\bibfnamefont{A.}~\bibnamefont{Perrier}},\ and\ \bibinfo {author}
  {\bibfnamefont{G.~J.}\ \bibnamefont{Kroes}},\ }%
  \bibfield{journal}{%
  \bibinfo {journal} {Chem. Phys. Lett.}\ }%
  \textbf{\bibinfo {volume} {434}},\ \bibinfo {pages} {231 } (\bibinfo {year}
  {2007})\BibitemShut{NoStop}%
\bibitem{behler1}%
  \BibitemOpen
  \bibfield{author}{%
  \bibinfo {author} {\bibfnamefont{J.}~\bibnamefont{Behler}}, \bibinfo {author}
  {\bibfnamefont{B.}~\bibnamefont{Delley}}, \bibinfo {author}
  {\bibfnamefont{S.}~\bibnamefont{Lorenz}}, \bibinfo {author}
  {\bibfnamefont{K.}~\bibnamefont{Reuter}},\ and\ \bibinfo {author}
  {\bibfnamefont{M.}~\bibnamefont{Scheffler}},\ }%
  \bibfield{journal}{%
  \Doi{10.1103/PhysRevLett.94.036104}{\bibinfo {journal} {Phys. Rev. Lett.}}\
  }%
  \textbf{\bibinfo {volume} {94}},\ \bibinfo {pages} {036104} (\bibinfo {month}
  {Jan}\ \bibinfo {year} {2005})\BibitemShut{NoStop}%
\bibitem{behler2}%
  \BibitemOpen
  \bibfield{author}{%
  \bibinfo {author} {\bibfnamefont{J.}~\bibnamefont{Behler}}, \bibinfo {author}
  {\bibfnamefont{K.}~\bibnamefont{Reuter}},\ and\ \bibinfo {author}
  {\bibfnamefont{M.}~\bibnamefont{Scheffler}},\ }%
  \bibfield{journal}{%
  \Doi{10.1103/PhysRevB.77.115421}{\bibinfo {journal} {Phys. Rev. B}}\ }%
  \textbf{\bibinfo {volume} {77}},\ \bibinfo {pages} {115421} (\bibinfo {month}
  {Mar}\ \bibinfo {year} {2008})\BibitemShut{NoStop}%
\bibitem{metiu}%
  \BibitemOpen
  \bibfield{author}{%
  \bibinfo {author} {\bibfnamefont{H.}~\bibnamefont{Metiu}}\ and\ \bibinfo
  {author} {\bibfnamefont{G.}~\bibnamefont{Sch{\"o}n}},\ }%
  \bibfield{journal}{%
  \bibinfo {journal} {Phys. Rev. Lett.}\ }%
  \textbf{\bibinfo {volume} {53}},\ \bibinfo {pages} {13} (\bibinfo {year}
  {1984})\BibitemShut{NoStop}%
\bibitem{schmid}%
  \BibitemOpen
  \bibfield{author}{%
  \bibinfo {author} {\bibfnamefont{A.}~\bibnamefont{Schmid}},\ }%
  \bibfield{journal}{%
  \bibinfo {journal} {J. Low Temp. Phys.}\ }%
  \textbf{\bibinfo {volume} {49}},\ \bibinfo {pages} {609} (\bibinfo {year}
  {1982})\BibitemShut{NoStop}%
\bibitem{caldeira}%
  \BibitemOpen
  \bibfield{author}{%
  \bibinfo {author} {\bibfnamefont{A.~O.}\ \bibnamefont{Caldeira}}\ and\
  \bibinfo {author} {\bibfnamefont{A.~J.}\ \bibnamefont{Leggett}},\ }%
  \bibfield{journal}{%
  \bibinfo {journal} {Physica A.}\ }%
  \textbf{\bibinfo {volume} {121}},\ \bibinfo {pages} {587} (\bibinfo {year}
  {1983})\BibitemShut{NoStop}%
\bibitem{brandbyge}%
  \BibitemOpen
  \bibfield{author}{%
  \bibinfo {author} {\bibfnamefont{M.}~\bibnamefont{Brandbyge}}, \bibinfo
  {author} {\bibfnamefont{P.}~\bibnamefont{Hedeg{\aa}rd}}, \bibinfo {author}
  {\bibfnamefont{T.~F.}\ \bibnamefont{Heinz}}, \bibinfo {author}
  {\bibfnamefont{J.~A.}\ \bibnamefont{Misewich}},\ and\ \bibinfo {author}
  {\bibfnamefont{D.~M.}\ \bibnamefont{Newns}},\ }%
  \bibfield{journal}{%
  \bibinfo {journal} {Phys. Rev. B}\ }%
  \textbf{\bibinfo {volume} {52}},\ \bibinfo {pages} {6042} (\bibinfo {year}
  {1995})\BibitemShut{NoStop}%
\bibitem{tully}%
  \BibitemOpen
  \bibfield{author}{%
  \bibinfo {author} {\bibfnamefont{J.~C.}\ \bibnamefont{Tully}}, \bibinfo
  {author} {\bibfnamefont{M.}~\bibnamefont{Gomez}},\ and\ \bibinfo {author}
  {\bibfnamefont{M.}~\bibnamefont{Head-Gordon}},\ }%
  \bibfield{journal}{%
  \bibinfo {journal} {J. Vac. Sci. Technol. A}\ }%
  \textbf{\bibinfo {volume} {11}},\ \bibinfo {pages} {1914} (\bibinfo {year}
  {1993})\BibitemShut{NoStop}%
\bibitem{luntz2}%
  \BibitemOpen
  \bibfield{author}{%
  \bibinfo {author} {\bibfnamefont{A.~C.}\ \bibnamefont{Luntz}}, \bibinfo
  {author} {\bibfnamefont{M.}~\bibnamefont{Persson}}, \bibinfo {author}
  {\bibfnamefont{S.}~\bibnamefont{Wagner}}, \bibinfo {author}
  {\bibfnamefont{C.}~\bibnamefont{Frischkorn}},\ and\ \bibinfo {author}
  {\bibfnamefont{M.}~\bibnamefont{Wolf}},\ }%
  \bibfield{journal}{%
  \bibinfo {journal} {J. Chem. Phys.}\ }%
  \textbf{\bibinfo {volume} {124}},\ \bibinfo {pages} {244702} (\bibinfo {year}
  {2006})\BibitemShut{NoStop}%
\bibitem{olsen5}%
  \BibitemOpen
  \bibfield{author}{%
  \bibinfo {author} {\bibfnamefont{T.}~\bibnamefont{Olsen}}\ and\ \bibinfo
  {author} {\bibfnamefont{J.}~\bibnamefont{Schi{\o}tz}},\ }%
  \bibfield{journal}{%
  \bibinfo {journal} {Submitted}\ }%
  \bibinfo {note} {, arxiv:1003.2318}\BibitemShut{NoStop}%
\bibitem{trail}%
  \BibitemOpen
  \bibfield{author}{%
  \bibinfo {author} {\bibfnamefont{J.~R.}\ \bibnamefont{Trail}}, \bibinfo
  {author} {\bibfnamefont{D.~M.}\ \bibnamefont{Bird}}, \bibinfo {author}
  {\bibfnamefont{M.}~\bibnamefont{Persson}},\ and\ \bibinfo {author}
  {\bibfnamefont{S.}~\bibnamefont{Holloway}},\ }%
  \bibfield{journal}{%
  \bibinfo {journal} {J. Chem. Phys.}\ }%
  \textbf{\bibinfo {volume} {119}},\ \bibinfo {pages} {4539} (\bibinfo {year}
  {2003})\BibitemShut{NoStop}%
\bibitem{newnsanderson1}%
  \BibitemOpen
  \bibfield{author}{%
  \bibinfo {author} {\bibfnamefont{P.~W.}\ \bibnamefont{Anderson}},\ }%
  \bibfield{journal}{%
  \bibinfo {journal} {Phys. Rev.}\ }%
  \textbf{\bibinfo {volume} {124}},\ \bibinfo {pages} {41} (\bibinfo {year}
  {1961})\BibitemShut{NoStop}%
\bibitem{newnsanderson2}%
  \BibitemOpen
  \bibfield{author}{%
  \bibinfo {author} {\bibfnamefont{D.~M.}\ \bibnamefont{Newns}},\ }%
  \bibfield{journal}{%
  \bibinfo {journal} {Phys. Rev.}\ }%
  \textbf{\bibinfo {volume} {178}},\ \bibinfo {pages} {1123} (\bibinfo {year}
  {1969})\BibitemShut{NoStop}%
\bibitem{gao97}%
  \BibitemOpen
  \bibfield{author}{%
  \bibinfo {author} {\bibfnamefont{S.}~\bibnamefont{Gao}},\ }%
  \bibfield{journal}{%
  \bibinfo {journal} {Phys. Rev. B}\ }%
  \textbf{\bibinfo {volume} {55}},\ \bibinfo {pages} {1876} (\bibinfo {year}
  {1997})\BibitemShut{NoStop}%
\bibitem{olsen1}%
  \BibitemOpen
  \bibfield{author}{%
  \bibinfo {author} {\bibfnamefont{T.}~\bibnamefont{Olsen}}, \bibinfo {author}
  {\bibfnamefont{J.}~\bibnamefont{Gavnholt}},\ and\ \bibinfo {author}
  {\bibfnamefont{J.}~\bibnamefont{Schi{\o}tz}},\ }%
  \bibfield{journal}{%
  \bibinfo {journal} {Phys. Rev. B}\ }%
  \textbf{\bibinfo {volume} {79}},\ \bibinfo {pages} {035403} (\bibinfo {year}
  {2009})\BibitemShut{NoStop}%
\bibitem{olsen3}%
  \BibitemOpen
  \bibfield{author}{%
  \bibinfo {author} {\bibfnamefont{T.}~\bibnamefont{Olsen}}\ and\ \bibinfo
  {author} {\bibfnamefont{J.}~\bibnamefont{Schi{\o}tz}},\ }%
  \bibfield{journal}{%
  \bibinfo {journal} {Phys. Rev. Lett.}\ }%
  \textbf{\bibinfo {volume} {103}},\ \bibinfo {pages} {238301} (\bibinfo {year}
  {2009})\BibitemShut{NoStop}%
\bibitem{feynman-vernon}%
  \BibitemOpen
  \bibfield{author}{%
  \bibinfo {author} {\bibfnamefont{R.~P.}\ \bibnamefont{Feynman}}\ and\
  \bibinfo {author} {\bibfnamefont{F.~L.}\ \bibnamefont{Vernon}},\ }%
  \bibfield{journal}{%
  \bibinfo {journal} {Ann. Phys. (N.Y.)}\ }%
  \textbf{\bibinfo {volume} {24}},\ \bibinfo {pages} {118} (\bibinfo {year}
  {1963})\BibitemShut{NoStop}%
\bibitem{olsen2}%
  \BibitemOpen
  \bibfield{author}{%
  \bibinfo {author} {\bibfnamefont{T.}~\bibnamefont{Olsen}},\ }%
  \bibfield{journal}{%
  \bibinfo {journal} {Phys. Rev. B}\ }%
  \textbf{\bibinfo {volume} {79}},\ \bibinfo {pages} {235414} (\bibinfo {year}
  {2009})\BibitemShut{NoStop}%
\bibitem{gpaw}%
  \BibitemOpen
  \bibinfo {note} {The \texttt{gpaw} code is available as a part of the CAMPOS
  software: \texttt{www.camd.dtu.dk/Software}}\BibitemShut{NoStop}%
\bibitem{mortensen}%
  \BibitemOpen
  \bibfield{author}{%
  \bibinfo {author} {\bibfnamefont{J.~J.}\ \bibnamefont{Mortensen}}, \bibinfo
  {author} {\bibfnamefont{L.~B.}\ \bibnamefont{Hansen}},\ and\ \bibinfo
  {author} {\bibfnamefont{K.~W.}\ \bibnamefont{Jacobsen}},\ }%
  \bibfield{journal}{%
  \bibinfo {journal} {Phys. Rev. B}\ }%
  \textbf{\bibinfo {volume} {71}},\ \bibinfo {pages} {035109} (\bibinfo {year}
  {2005})\BibitemShut{NoStop}%
\bibitem{blochl1}%
  \BibitemOpen
  \bibfield{author}{%
  \bibinfo {author} {\bibfnamefont{P.~E.}\ \bibnamefont{Bl{\"o}chl}},\ }%
  \bibfield{journal}{%
  \bibinfo {journal} {Phys. Rev. B}\ }%
  \textbf{\bibinfo {volume} {50}},\ \bibinfo {pages} {17953} (\bibinfo {year}
  {1994})\BibitemShut{NoStop}%
\bibitem{blochl2}%
  \BibitemOpen
  \bibfield{author}{%
  \bibinfo {author} {\bibfnamefont{P.~E.}\ \bibnamefont{Bl{\"o}chl}}, \bibinfo
  {author} {\bibfnamefont{C.~J.}\ \bibnamefont{F{\"o}rst}},\ and\ \bibinfo
  {author} {\bibfnamefont{J.}~\bibnamefont{Schimpl}},\ }%
  \bibfield{journal}{%
  \bibinfo {journal} {Bull. Mat. Sci.}\ }%
  \textbf{\bibinfo {volume} {26}},\ \bibinfo {pages} {33} (\bibinfo {year}
  {2003})\BibitemShut{NoStop}%
\bibitem{hammer}%
  \BibitemOpen
  \bibfield{author}{%
  \bibinfo {author} {\bibfnamefont{B.}~\bibnamefont{Hammer}}, \bibinfo {author}
  {\bibfnamefont{L.~B.}\ \bibnamefont{Hansen}},\ and\ \bibinfo {author}
  {\bibfnamefont{J.~K.}\ \bibnamefont{N{\o}rskov}},\ }%
  \bibfield{journal}{%
  \bibinfo {journal} {Phys. Rev. B}\ }%
  \textbf{\bibinfo {volume} {59}},\ \bibinfo {pages} {7413} (\bibinfo {year}
  {1999})\BibitemShut{NoStop}%
\bibitem{gavnholt}%
  \BibitemOpen
  \bibfield{author}{%
  \bibinfo {author} {\bibfnamefont{J.}~\bibnamefont{Gavnholt}}, \bibinfo
  {author} {\bibfnamefont{T.}~\bibnamefont{Olsen}}, \bibinfo {author}
  {\bibfnamefont{M.}~\bibnamefont{Engelund}},\ and\ \bibinfo {author}
  {\bibfnamefont{J.}~\bibnamefont{Schi{\o}tz}},\ }%
  \bibfield{journal}{%
  \bibinfo {journal} {Phys. Rev. B}\ }%
  \textbf{\bibinfo {volume} {78}},\ \bibinfo {pages} {075441} (\bibinfo {year}
  {2008})\BibitemShut{NoStop}%
\end{thebibliography}

%

\end{document}